\documentclass[
    prl, twocolumn, superscriptaddress, nofootinbib, amsmath, amssymb,
    aps, floatfix, preprintnumbers
]{revtex4-2}
\usepackage[utf8]{inputenc}
\usepackage{setspace}
\usepackage[dvips]{graphicx}
\usepackage{booktabs}
\usepackage[dvipsnames]{xcolor}
\usepackage{tikz}
\definecolor{CiteBlue}{RGB}{45,52,151}
\usepackage[
    colorlinks=true,
    linkcolor=CiteBlue,
    urlcolor=CiteBlue,
    citecolor=CiteBlue
]{hyperref}
\usepackage[version=4]{mhchem}
\usepackage{aas_macros}

\usepackage[capitalise]{cleveref}
\usepackage{siunitx}
\DeclareSIUnit{\year}{yr}
\usepackage{physics}

\newcommand{\nn}{\nonumber}

\usepackage{bm}

\newsavebox{\twosubbox}

\begin{document}

\title{Probing Dark Matter--Electron Interactions with Superconducting Qubits}

\author{Yonit Hochberg}
\affiliation{Racah Institute of Physics, Hebrew University of Jerusalem, Jerusalem 91904, Israel}
\affiliation{Laboratory for Elementary Particle Physics,
 Cornell University, Ithaca, NY 14853, USA}

\author{Majed Khalaf}
\affiliation{Racah Institute of Physics, Hebrew University of Jerusalem, Jerusalem 91904, Israel}
\affiliation{Laboratory for Elementary Particle Physics,
 Cornell University, Ithaca, NY 14853, USA}
 
\author{Noah Kurinsky}
\affiliation{SLAC National Accelerator Laboratory, 2575 Sand Hill Rd, Menlo Park, CA 94025, USA}
\affiliation{Kavli Institute for Particle Astrophysics and Cosmology, Stanford University, Stanford, CA 94035, USA}

\author{Alessandro Lenoci}
\affiliation{Racah Institute of Physics, Hebrew University of Jerusalem, Jerusalem 91904, Israel}
\affiliation{Laboratory for Elementary Particle Physics,
 Cornell University, Ithaca, NY 14853, USA}

\author{Rotem Ovadia}
\affiliation{Racah Institute of Physics, Hebrew University of Jerusalem, Jerusalem 91904, Israel}
\affiliation{Laboratory for Elementary Particle Physics,
 Cornell University, Ithaca, NY 14853, USA}

\date\today

\begin{abstract}\ignorespaces{}
    Quantum device measurements are powerful tools to probe dark matter interactions. 
    Among these, transmon qubits stand out for their ability to suppress external noise while remaining highly sensitive to tiny energy deposits. 
    Ambient galactic halo dark matter interacting with electrons can deposit energy in the qubit, leading to changes in its decoherence time.
    Recent measurements of transmons have consistently measured, in various experimental setups, a residual contribution to the decoherence time unexplained by thermal noise or known external sources.
    We use such measurements to set the most stringent laboratory-based constraints to date on dark matter–electron scattering at the keV scale and competitive constraints on dark photon absorption.
\end{abstract}

\maketitle

\section{Introduction}
Despite decades of impressive experimental efforts in detection of dark matter (DM), its fundamental nature is still unknown. Null observation of DM at the electroweak scale has brought sub-GeV DM into the limelight, where many new experiments sensitive to small energy deposits are being devised~\cite{Essig:2022dfa,SENSEI:2023zdf,Hochberg:2019cyy,Hochberg:2021yud,Gao:2024irf,Abbamonte:2025guf}. 
Among these, superconducting targets stand out with their sensitivity to energy deposits of ${\cal O}({\rm meV})$, where DM interactions could break Cooper pairs in the system, resulting in a readable signal~\cite{Hochberg:2015pha,Hochberg:2015fth}. Indeed, several superconducting technologies have already demonstrated their ability to probe light DM in the sub-MeV mass range interacting with electrons, including superconducting nanowire single photon detectors~(SNSPDs)~\cite{Hochberg:2019cyy,Hochberg:2021yud,QROCODILE:2024zmg}, kinetic inductance devices~(KIDs)~\cite{Gao:2024irf}, and transition edge sensors~(TESs)~\cite{Schwemmbauer:2024rcr}. Experimental progress along these lines is enabled thanks to the rapid advancement of quantum sensing technologies and the application of such advancements towards fundamental physics~\cite{Buchmueller:2022djy,Baxter:2025odk}. 

In this work, we focus on transmon qubits~\cite{Koch:2007hay,Blais:2020wjs}, a type of superconducting qubit characterized by incredible 
sensitivity to background noises.
Coherence measurements of transmons provide a sensitive probe of power injection from external sources. 
The relevant energy scale in these systems is the Cooper-pair binding energy of ${\cal O}({\rm meV})$, which sets the threshold for energy deposits of interest. 
Transmons, through their sensitivity to changes in the local quasiparticle environment, are thus sensitive to DM with mass as low as the keV scale that scatters with electrons in the qubits, depositing its kinetic energy into the system~\cite{Hochberg:2021yud, Hochberg:2021ymx}, as well as to the absorption of dark photon DM with meV-scale masses, where the entire mass-energy of the DM is deposited into the system~\cite{Hochberg:2016ajh, Knapen:2021run}.

Broken Cooper pairs result in charged quasiparticle~(QP) excitations in the superconductor.
The existence of QPs well in excess of those expected to be produced thermally is often referred to as QP poisoning, and has been observed in a variety of superconducting devices~\cite{deVisser:2011ar} including resonators~\cite{Zmuidzinas:2012rev, deVisser:2014qp}, KIDs~\cite{Mazin:2009rev}, and transmons~\cite{Riste:2013zqw,Diamond:2022scj,Connolly:2023gww}.
In some implementations, QP poisoning is asymptotically curbed by QP recombination at sufficiently large  populations~\cite{deVisser:2014qp,Goldie:2012qp}.
However, in transmons, even a residual QP fraction as small as $x_{\rm qp} \sim 10^{-9}$ can degrade coherence~\cite{Riste:2013zqw,Diamond:2022scj,Connolly:2023gww}, rendering QP poisoning a limiting factor.
Indeed, the mitigation of QP poisoning has become a central focus of ongoing research on superconducting qubits, in pursuit of surpassing the physical error rate threshold for scalable fault-tolerant quantum computing~\cite{Aharonov:1999ei}.

The microscopic origin of the experimentally observed QP excess in transmons, and in fact most superconducting devices, remains unknown, with proposed explanations including stray infrared or microwave photons, natural radioactivity, and cosmic-ray interactions~\cite{Wilen:2020lgg, Iaia:2022exc, McEwen:2021wdg, Andreev:2021exc, Cardani:2020vvp, Mannila:2021exc, Bespalov2016, Vepsalainen:2020trd, Bargerbos:2023exc}. Refs.~\cite{Connolly:2023gww,Pan:2022kpz,Liu:2022aez} illustrate ongoing efforts to understand the origin of residual QPs and to mitigate their occurrence through improved shielding, refined background modeling, and circuit modifications aimed at reducing the flux of Cooper-pair–breaking photons.
An intriguing possibility is that part or all of this excess arises from interactions between galactic DM and the device, though with the current state-of-the-art modeling of transmon dynamics, the DM signal remains indistinguishable from other backgrounds.
One can, however, use these measurements to place world leading bounds on DM interactions by requiring the total DM power deposit does not exceed the observed one.

Leveraging the intrinsic sensitivity of power measurements in transmons~\cite{Riste:2013zqw}, recent studies demonstrated that existing data from these devices can be used to place constraints on DM scattering with nucleons in the target~\cite{Das:2022srn,Das:2024jdz} and on light bosons absorbed by the target's electrons~\cite{Chao:2024owf}. 
Here we provide a complementary and revised analysis using recent state-of-the-art measurements of transmons~\cite{Connolly:2023gww}, where additional measures were employed to shield the transmon from external radiation. We use this data to place the first constraints on {\it DM-electron} scattering from qubit transmons, where our new limits surpass all existing terrestrial constraints on DM masses $\sim 1$--$100\ {\rm keV}$. We also provide updated dark photon absorption limits and projections for future reduction of residual QP population in these devices.

This {\it Letter} is organized as follows. We begin by reviewing how DM interactions with electrons relate to noise measurements in transmons. We  describe our modeling of the qubit response and discuss relevant transmon data~\cite{Riste:2013zqw, Connolly:2023gww}. We then use this existing data to place novel limits on DM-electron interactions, including the strongest terrestrial constraints to date on DM-electron scatterings down to the keV scale. 
Throughout, we use natural units where $\hbar=c=k_{\rm B}=1$.

\section{DM-induced noise in transmons}
Physical two-level systems suffer from both dephasing and relaxation, where the relative phase between their two levels becomes uncertain due to noise or environmental fluctuations (\textit{e.g.} magnetic noise, charge noise, or phonons \cite{Siddiqi21}), which results in a finite coherence time. This is the case for transmons, which are the main form of superconducting qubits used in contemporary quantum computing. A key component of the transmon is the Josephson junction, made of two superconducting electrodes connected by an insulating thin barrier through which Cooper pairs can tunnel.  
Importantly, the energy splitting of the transmon depends on its overall charge parity.
Since the charge of a single cooper pair is $2e$, the tunneling of Cooper pairs across the junction does not alter the charge parity of the transmon, thereby preserving the coherence of the qubit.

When Cooper pairs break, however, the resulting QPs may also tunnel through the Josephson junction, flipping the charge parity and contributing to  dephasing and relaxation that limits the coherence time.
Monitoring of the coherence time thus offers a method to probe gap-scale energy depositions in the transmon. In particular, DM interacting weakly with electrons can penetrate the transmon and deposit its energy in the device's material. 
Such energy deposits exceeding the Cooper pair binding energy of $2 \, \Delta$ would appear as a residual QP density in the transmon.

In this work, we consider energy deposits in the Josephson junction and in the thin aluminum films that constitute the transmon.
Adopting the notation ubiquitous in recent literature, we define the fraction of QPs in the qubit as $x_{\rm qp} = n_{\rm qp} / (2 \Delta \nu_0)$ where $n_{\rm qp}$ is the QP number density, $\Delta \simeq 170\ \mu{\rm eV}$ is the Cooper pair binding energy in Al, and $\nu_0 = 1.2 \times 10^{4}\ \mu{\rm {\rm m}^{-3}} {\mu}{\rm eV}^{-1}$ is the Cooper pair density of states at the Fermi level in Al.
 In the mean-field approximation, one can write a master equation for the QP fraction $x_{\rm qp}$ as~\cite{Wang:2014hnf, Diamond:2022scj}
\begin{equation}\label{eq:qp master equation}
    \dot{x}_{\rm qp} = \Gamma_{\rm G} - r \, x_{\rm qp}^2 - s \, x_{\rm qp}\ ,
\end{equation}
where $\Gamma_{\rm G}$, $r\, x_{\rm qp}^2$, and $s\,  x_{\rm qp}$ are the quasiparticle generation, recombination, and trapping rates, respectively. 
The generation rate accounts for QPs formed by external energy-injection ({\it e.g.} DM) or thermal fluctuations.
The recombination rate accounts for the ability of two quasiparticles to recombine into a Cooper pair by emitting a phonon, with $r$ a material constant.
The trapping rate accounts for trapping of QPs in regions characterized by a reduced superconducting gap, with $s$ a system parameter characterizing the abundance of such regions.
Considering the steady-state solution $\dot x_{\rm qp}\simeq 0$, Eq.~\eqref{eq:qp master equation} becomes 
\begin{equation}\label{eq:qp_equilibrium}
    \Gamma_{\rm G} = r \, x_{\rm qp}^2 + s \, x_{\rm qp} \, .
\end{equation}
The relevant term on the right hand side is determined by the experimental setup, which ultimately determines the values of $r, s$.

To determine if a signal is generated by external sources, the number of excess QPs should be compared to their expected thermal population. 
Following Ref.~\cite{lutchyn2005quasiparticle,lutchyn2006PRB,Koch:2007hay}, thermal fluctuations at temperature $T$ produce a temperature-dependent quasiparticle density
\begin{align}
    x_{\rm qp}^{\rm th}(T)  = \sqrt{\frac{2 \pi T}{ \Delta }}\exp(-\Delta/T)\, .
\end{align}
For superconducting Al, $x^{\rm th}_{\rm qp}(88\, {\rm mK})= 10^{-10}$ and $x^{\rm th}_{\rm qp}(110 \, {\rm mK}) = 10^{-8}$. 
At temperatures where $x_{\rm qp}^{\rm th}(T) \ll x_{\rm qp}$ the thermal population can be neglected.
Both measurements~\cite{Riste:2013zqw,Connolly:2023gww} we consider in this work are in a regime where the thermal population is negligible; we thus omit the thermal component in what follows. 
Following Ref.~\cite{Das:2024jdz}, the QP generation rate in the low temperature regime is related to the power injection as
\begin{equation}\label{eq:GammaG}
    \Gamma_{\rm G} \geq \frac{\rho_{\rm T}}{2 \Delta^2 \nu_0} \int d\omega \ \omega \frac{d\Gamma}{d\omega} = \frac{D_{\rm dm}}{2 \Delta^2 \nu_0}\ ,
\end{equation}
where $\rho_{\rm T}=2.7\ {\rm g/cm^3}$ for Al, $\omega$ is the energy deposited in the target by an incident particle (such as DM), and $d\Gamma/d\omega$ is the spectral interaction rate per unit exposure.
The definition of $D_{\rm dm}$ follows directly from Eq.~\eqref{eq:GammaG} with $\Gamma$ taken as the DM-electron scattering rate per unit volume. 
The inequality indicates that the DM energy deposit in the transmon is an irreducible contribution to the QP generation rate.

To compute $D_{\rm dm}$, we use the relation between the spectral rate arising from DM-electron elastic scattering and the longitudinal dielectric function $\epsilon_L(\omega, \vb{q})$ of the target, which is given by~\cite{Hochberg:2021pkt, Hochberg:2025rjs}
\begin{align}\label{eq:spectral rate}
    \frac{d \Gamma}{d\omega} = & \frac{\pi \bar \sigma_{e} \rho_{\chi}}{\rho_{\rm T} m_{\chi} \mu_{e \chi}^2}
    \int d^3 {\bf v} \, f_{\chi}({\bf v})\\\nonumber 
    &
    \times \int \frac{d^3 {\bf q}}{(2\pi)^3} \mathcal{F}^2(q)\, \frac{q^2}{2\pi \alpha} \, {\rm Im} \pqty{ -\frac{1}{\epsilon_L(\omega,{\bf q})} } \, \delta (\omega - \omega_{\bf q}) \ ,
\end{align}
with ${\rm Im} \pqty{ -1/\epsilon_L(\omega,{\bf q}) }$ the loss function of the material. 
Here $m_\chi$ is the DM mass, $\mu_{e\chi}$ is the DM--electron reduced mass, ${\bf q}$ is the transferred momentum with $q = |{\bf q}|$, and $\bar{\sigma}_e = (\mu_{e\chi}^2/\pi)\, |V(q_0)|^2$ is the fiducial cross section with $q_0 = \alpha m_e$ a reference momentum with $\alpha$ the electromagnetic fine structure constant and $m_e$ the electron mass.
The DM--electron interaction is taken to be a Yukawa-type potential $V(q) = g_{e\chi}^2(q^2 + m_{\phi}^2)^{-1}$, where $g_{e\chi}$ is the effective coupling between DM and electrons and $m_{\phi}$ is the mass of the mediator. 
The ambient DM density is taken as $\rho_\chi \simeq 0.4~{\rm GeV}/{\rm cm^3}$ and the DM velocity distribution $f_\chi({\bf v})$ is given by the Standard Halo Model~\cite{Lewin:1995rx} with $v_0 = 220 \, {\rm km}/{\rm s}$, $v_{\oplus} = 232\, {\rm km}/{\rm s}$, and $v_{\rm esc} = 540 \, {\rm km}/{\rm s}$.
The form factor is given by $\mathcal{F}^2(q) = |V(q)|^2 / |V(q_0)|^2$ which is $1$ for a heavy mediator and $(q_0/q)^{4}$ for a light mediator. 
The transferred energy in the scattering process is determined by kinematics, $\omega_{\bf q} = {\bf q} \cdot {\bf v} - {q^2}/{2 m_\chi}$.

transmons are also sensitive to absorption of DM by electrons. 
As a benchmark, we consider dark photon  DM $A'$ that is kinetically mixed with the standard photon. In this case, the power density deposited by DM is given by \cite{Hochberg:2019cyy, Knapen:2021bwg, Chao:2024owf}
\begin{equation}\label{eq:abs}
    D_{\rm dm} 
    =  \kappa^2 m_{A'} \rho_{ A'}{\rm Im} \pqty{-\frac{1}{\epsilon_L(m_{A'},{\bf q} \to 0)}} \ .
\end{equation}
Here ${\kappa}$ is the kinetic mixing coupling, $m_{A'}$ the dark photon mass, $\rho_{A'}\simeq 0.4\, {\rm GeV/cm^3}$ is the ambient DM density.
The deposited power in absorption   Eq.~\eqref{eq:abs} is determined by the loss function at effectively zero momentum $ q \rightarrow 0$, since $ q \simeq 10^{-3} m_{A'}\ll m_{A'}$.

\section{Transmon response modeling}
%
To calculate the transmon’s response to external energy and momentum injection, we model it as a thin superconducting aluminum (Al) film. The Al response is described using an isotropic Lindhard function $\epsilon_{\rm Lind}$ with a plasmon frequency $\omega_p = 15.8\,{\rm eV}$ and width $\Gamma_p = 1.58\,{\rm eV}$. Following Refs.~\cite{Allen:1971coh,Hochberg:2016ajh}, we apply a superconducting coherence factor 
that accounts for the variation between ordinary conductors and superconductors near the superconducting gap $2 \, \Delta$.

\begin{figure}[t]
    \centering
    \includegraphics[width=.95\linewidth] {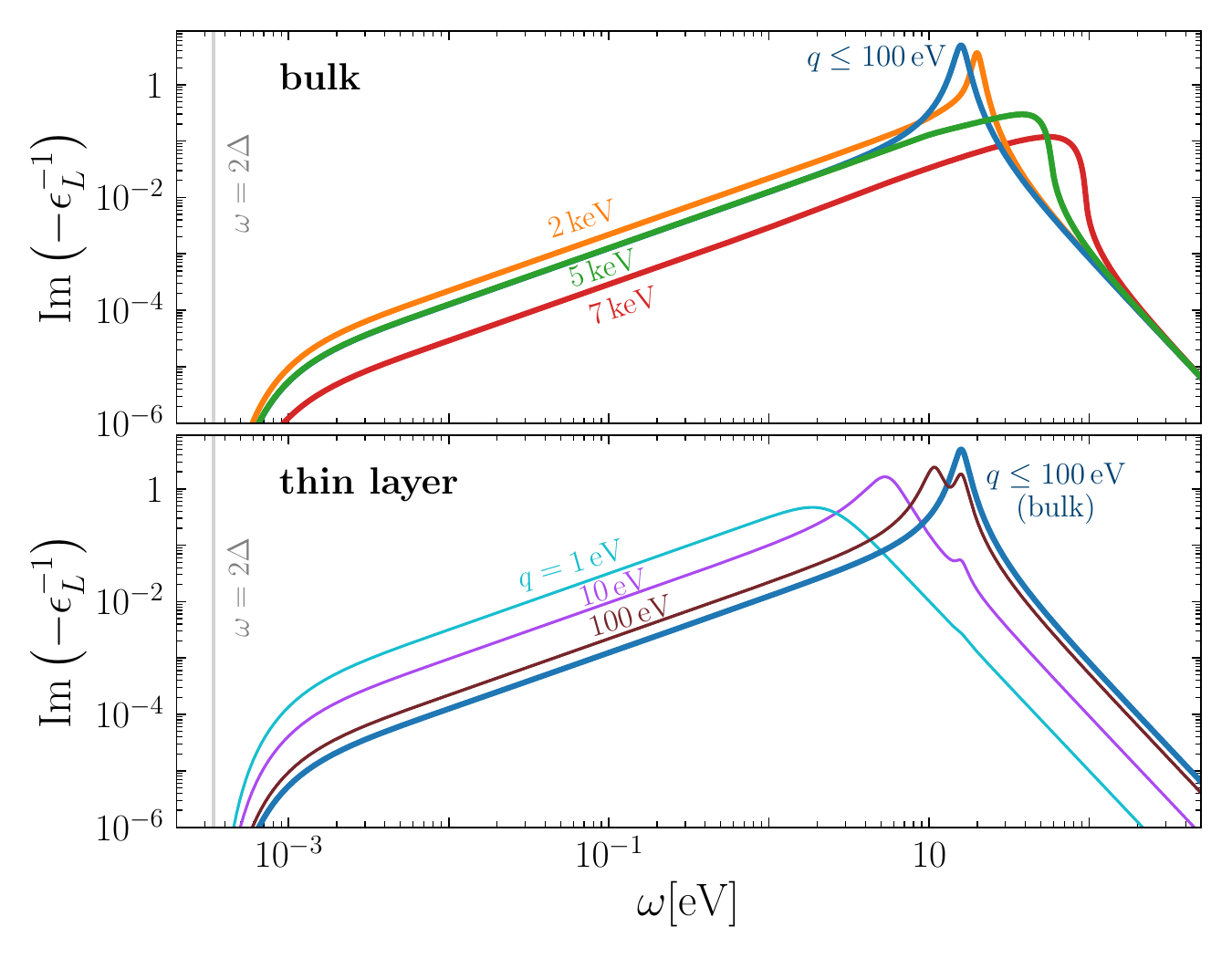}
    \caption{
    {\bf Al loss function.} 
    {\it Top:} The bulk Al-based transmon response. 
    We describe the overall response of Al by a Lindhard function, augmented by a near-superconducting-gap correction factor.
    For $q$ smaller than a few hundred eV, the loss is approximately momentum-independent.
    {\it Bottom:} The transmon loss function including momenta-dependent corrections due to the thin layer geometry, with momenta deposits taken parallel to the layer. 
    {\it Both panels:}~The thick (thin) solid colored curves delineate the bulk (or thin-layer) response for representative values of the momenta magnitude~$q$, and the gray line delineates the superconducting gap. 
    For $q \gtrsim 2\, {\rm keV} \gg d^{-1}$, with $d$ the Al layer thickness, there is no discernible difference between the bulk and thin-layer loss functions. 
    }
    \label{fig:Al loss}
\end{figure}

In the context of scattering, we consider the effect of the thin layer geometry of the transmon.
We model the transmon as an infinite $d \simeq 5 \, {\rm nm}$ thin layer of Aluminum placed between vacuum and an insulating Al$_2$O$_3$ substrate, and use the methodology described in Refs.~\cite{Lasenby:2021wsc, Hochberg:2021yud, QROCODILE:2024zmg} to find the thin-layer momenta-dependent modifications to the bulk loss function. (We note that this is only a proxy of the actual Josephson junction geometry which has a non-uniform width and an additional oxide layer within the Al). Similarly to Refs.~\cite{Hochberg:2021yud,QROCODILE:2024zmg} we only consider energy depositions in the Al layer, neglecting the absorptive part of the insulating substrate response, and allow for dissipation throughout. The sapphire substrate is approximated by a constant dielectric constant $\epsilon_{{\rm Al}_2{\rm O}_3} \simeq 3$~\cite{Djurii:1998}.
Fig.~\ref{fig:Al loss} presents the bulk and thin-layer loss functions for several representative values of momenta $q$. We find that the bulk loss function at fixed $\omega$ does not vary much for momenta $q \lesssim 1 \, {\rm keV}$, and that the thin layer geometry only affects the loss at $q \lesssim 2 \, {\rm keV}$.

In our calculations of the power density $D_{\rm dm}$ we apply an additional a momentum-independent factor to the loss function, corresponding to the quasiparticle production efficiency in thin-layer Al.
We take this factor according to the prescription given in Ref.~\cite{Guruswamy:2014qp},
\begin{eqnarray}
    f_{\text{eff}}(\omega)\approx \begin{cases} 
        1 -\frac{\omega - 2\, \Delta}{4\, \Delta} & 2 \Delta < \omega < 4 \Delta \\
        \frac{1}{2} & 4\Delta < \omega 
    \end{cases}\, .
\end{eqnarray}

\section{Extracting dark matter bounds}
Ref.~\cite{Connolly:2023gww} measured $x_{\rm qp} = 5.6 \times 10^{-10}$  
at $T = 20 \, {\rm mK}$, and reports estimates of $r \sim 10^{-2} \, {\rm GHz}$ and $s \sim 10^{1} - 10^{3} \, {\rm Hz}$ based on measurements of similar systems \cite{Wang:2014hnf, Diamond:2022scj}.
(We note that the values of $r$ and $s$ can vary between different experimental setups. 
For example, with different setups Refs.~\cite{Yelton:2024tqo, Yelton:2025wsy} measured $s \sim 10^{3}-10^4\, {\rm Hz}$.)
In the system of Ref.~\cite{Connolly:2023gww} the QP trapping rate is much larger than the recombination rate $r x_{\rm qp} \ll s$ and the thermal population of QPs is negligible, $x^{\rm th}_{\rm qp} \ll x_{\rm qp}$.
Based on this observation we can approximate from Eq.~\eqref{eq:qp_equilibrium} $  \Gamma_{\rm G} \simeq s x_{\rm qp}$ and use Eq.~\eqref{eq:GammaG} to  place an upper bound on the DM induced spectral density, 
\begin{eqnarray}\label{eq:trapping bound}
    D_{\rm dm} & \lesssim & 2 \, \Delta^2\,  \nu_0 \, s \,  x_{\rm qp} \\
    & \simeq & \pqty{6\times10^{-24} \, {\rm W} \,  \mu{\rm m}^{-3}} \pqty{\frac{x_{\rm qp}}{5.6 \times 10^{-10}}} \pqty{\frac{s}{10^2\, {\rm Hz}}} \, . \nn 
\end{eqnarray}

\begin{figure*}[ht!]
    \centering
    \includegraphics[width=.48\linewidth] {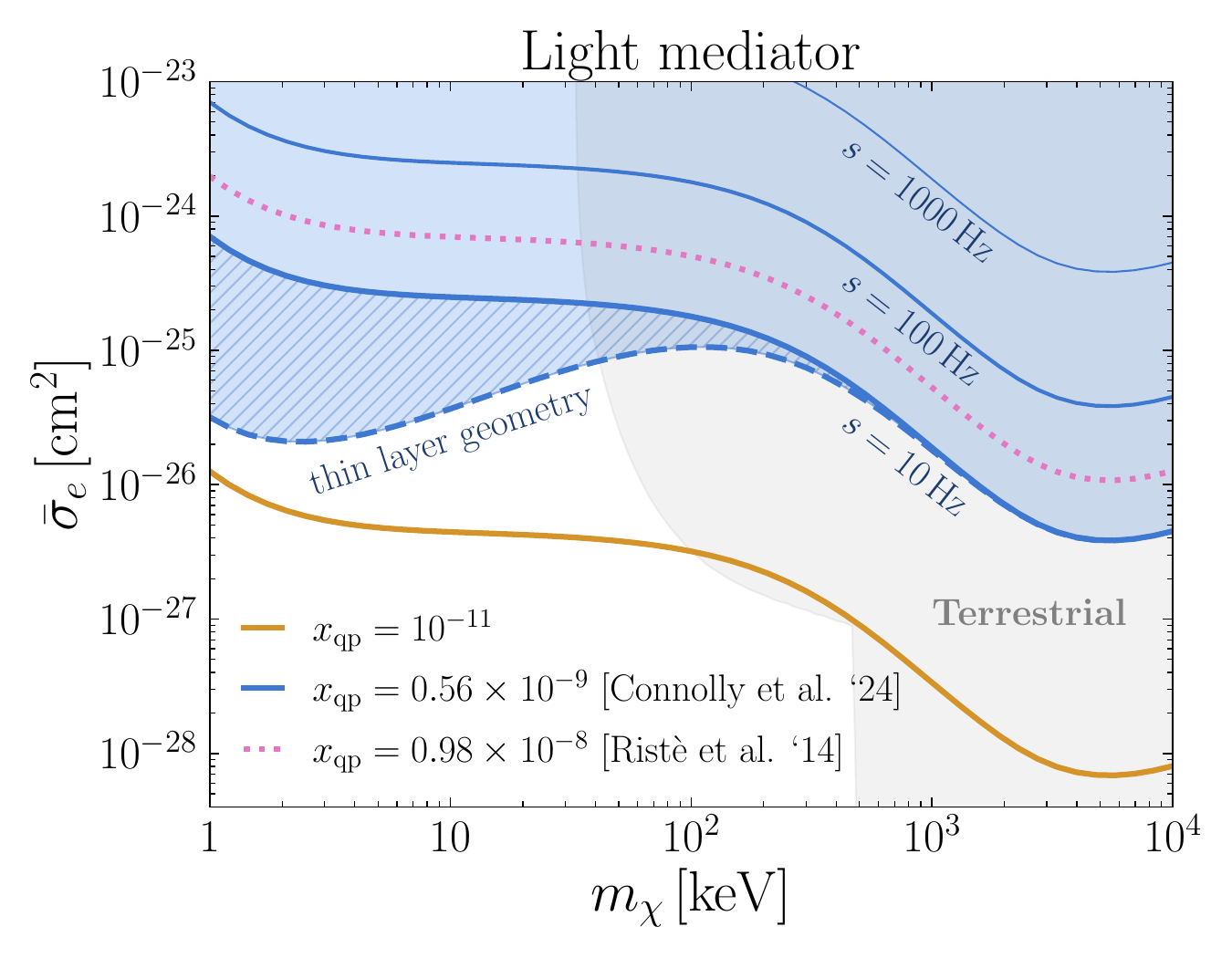}
    \includegraphics[width=.48\linewidth]{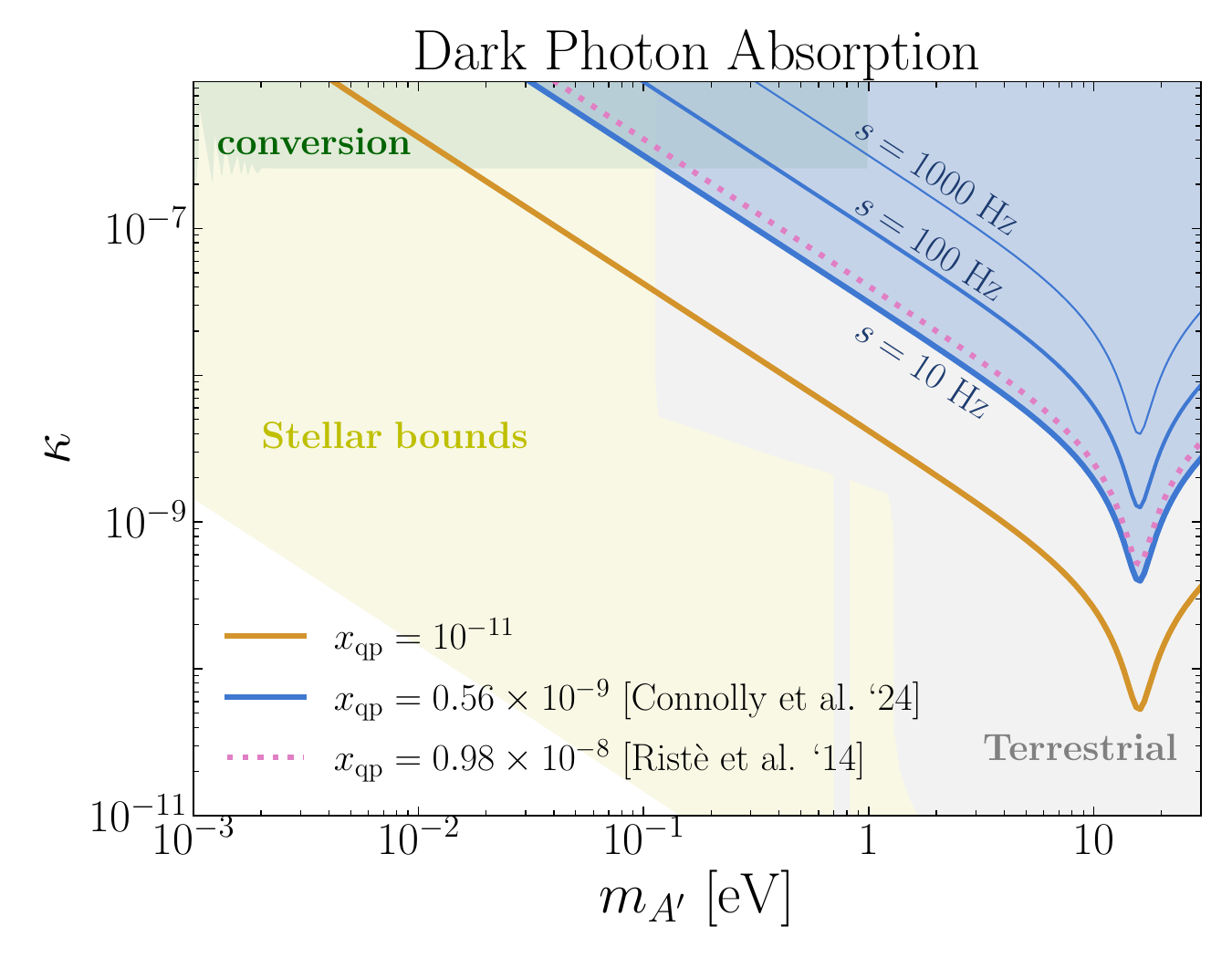}
    \caption{ 
{\bf DM-electron interaction constraints from transmons.}
{\it Left:} Constraints on DM-electron scattering. We show limits on the reference cross section for galactic halo DM interacting with electrons via a light mediator, as derived from transmon measurements. The results obtained from the data of Ref.~\cite{Connolly:2023gww} for different values of $s$ are shown as solid blue lines, while the limit placed via the measurement of Ref.~\cite{Riste:2013zqw} is indicated by the pink dotted line. 
The blue hatched region depicts the enhanced reach from the thin layer geometry (see text for  details). 
The projected sensitivity for a quantum device with $s = 10~{\rm Hz}$ and $x_{\rm qp} = 10^{-11}$ is shown as a solid orange line. The gray-shaded region represents existing experimental constraints~\cite{QROCODILE:2024zmg,SENSEI:2025qvp,DAMIC-M:2025luv}.  
Constraints may be relaxed for $\overline{\sigma}_e \gtrsim 10^{-23}\ {\rm cm}^2$ due to atmospheric scattering (see text).
{\it Right:} Constraints on dark photon DM absorption. We display upper bounds on the kinetic mixing parameter of dark photon DM obtained from the data of Ref.~\cite{Connolly:2023gww} for different values of $s$, shown as solid blue lines of varying thickness. The limit extracted from the measurement of Ref.~\cite{Riste:2013zqw} is shown  as an orange dotted line, and the projected sensitivity for a device with $x_{\rm qp} = 10^{-11}$ and $s = 10~{\rm Hz}$ is indicated by the dashed blue line. The gray-shaded region indicates existing terrestrial constraints assuming dark photon DM~\cite{EDELWEISS:2019vjv, EDELWEISS:2022ktt, DarkSide-50:2022qzh, Franco:2023sjx, SuperCDMS:2020aus, SuperCDMS:2023sql, CRESST:2019jnq, SENSEI:2023zdf, PandaX:2023xgl, LUX:2018akb, Essig:2019xkx, QROCODILE:2024zmg, Barak:2020fql, Amaral:2020ryn, Aguilar-Arevalo:2019wdi, Essig:2017kqs, Agnes:2018oej, XENON:2019gfn, An:2014twa, Agnese:2018col, Aguilar-Arevalo:2019wdi, Arnaud:2020svb, FUNKExperiment:2020ofv, Barak:2020fql}. The yellow-shaded region indicates complementary stellar constraints on dark photon absorption~\cite{An:2013yua, An:2020bxd}, while the green-shaded area represents laboratory bounds on photon–dark photon conversion~\cite{Bahre:2013ywa}.}\label{fig:bounds}
\end{figure*}

Following Refs.~\cite{Das:2022srn, Das:2024jdz}, we can also perform a complementary analysis of the older measurements of Ref.~\cite{Riste:2013zqw}.
Ref.~\cite{Riste:2013zqw} measured a residual steady state population of $x_{\rm qp} = (0.98 \pm 0.25) \times 10^{-8}$ in an Al based transmon at temperatures of $T=20 \, {\rm mK}$.
The values of $r,s$ of this experimental system were not reported, thus for our analysis we use the same estimates used in Refs.~\cite{Das:2022srn, Das:2024jdz} of $r = 0.1 \, {\rm GHz}$ and $s \ll 1 \, {\rm Hz}$, where the QP recombination rate is the dominant term in Eq.~\eqref{eq:qp_equilibrium}.    
Under these assumptions, $\Gamma_{\rm G} \simeq r x_{\rm qp}^2$ and the thermal population of QPs is negligible.
Similarly to before, we can place an upper bound on the DM induced spectral density from this measurement of 
\begin{eqnarray}\label{eq:recombination bound}
    D_{\rm dm} & \lesssim & 2 \, \Delta^2 \, \nu_0 \, x_{\rm qp}^2 \, r  \\ 
    & \simeq & \pqty{ 10^{-24}\ {\rm W}\ {\rm \mu {\rm m}^{-3}} } \pqty{\frac{r}{0.1 \, {\rm GHz}} } \pqty{\frac{x_{\rm qp}}{9.8 \times 10^{-9}}}^{2} \, . \nn
\end{eqnarray}
Note that in this case of the recombination regime, the power  deposition scales quadratically in $x_{\rm qp}$, as opposed to the trapping regime of Eq.~\eqref{eq:trapping bound} where the 
power deposition is linear in $x_{\rm qp}$.

It what follows, we use the inequalities Eqs.~\eqref{eq:trapping bound} and Eq.~\eqref{eq:recombination bound} along with Eqs.~\eqref{eq:GammaG}, \eqref{eq:spectral rate} and \eqref{eq:abs} to extract novel limits on light DM interactions with electrons from transmon data.

\section{Results}
%
The bounds from transmon measurements on DM-electron scattering with a light mediator are shown in the left panel of Fig.~\ref{fig:bounds}. 
The blue shaded region presents the bounds we extracted based on the residual noise measurements of Ref.~\cite{Connolly:2023gww}, using Eq.~\eqref{eq:trapping bound}, with different values of the trapping rate $s = 10\,,100\,, 1000\,{\rm Hz}$ delineated by progressively thinner solid lines.
The hatched blue region depicts the impact of the thin layer morphology, calculated by considering DM momenta deposits  
in various directions, resulting in an enhanced reach.  
We find that the maximal enhancement is achieved for momenta deposits parallel to the thin layer.
In the absence of knowing the direction of the DM wind at the time of measurement, the thin-layer result we show should be taken as a proxy to the true bound, which resides within the hatched region. 
Existing terrestrial bounds~\cite{QROCODILE:2024zmg,DAMIC-M:2025luv,SENSEI:2025qvp} are depicted by the shaded gray region.

We learn that the transmon measurement provides the strongest lab-based constraints to date on  DM-electron spin-independent interactions for DM masses below $30\,{\rm keV}$. 
For $s \lesssim 100 \, {\rm Hz}$, the bound extends below the estimated cross sections at which
atmospheric overburden becomes relevant, 
$\bar{\sigma}_e \lesssim 10^{-23} \, {\rm cm}^2$~\cite{Emken:2019tni}.
Given that $s=10 \, {\rm Hz}$ is reported in a recent measurement of a similar system~\cite{Diamond:2022scj}, we expect the true value of $s$ in the system considered here to fall in the lower end of the depicted range, corresponding to the stronger constraints.
The blue dashed curve delineates the projected reach for a future measurement of a residual fraction of QPs $x_{\rm qp} = 10^{-11}$ with $s=10 \, {\rm Hz}$.
The orange dashed curve delineates the bound we extract based on the older measurement of Ref.~\cite{Riste:2013zqw}, using Eq.~\eqref{eq:recombination bound}, complementary---for electronic interactions---to the nuclear scattering bound set in Ref.~\cite{Das:2024jdz}.
We find that, given the estimated fiducial value of $r \simeq 0.1 \, {\rm GHz}$, this measurement provides a slightly weaker bound than the one obtained from the measurement in Ref.~\cite{Connolly:2023gww}. 
(The projected reach for DM-electron scattering via a heavy mediator sits within regions already excluded by existing terrestrial constraints, and hence omit it from further discussion.)

Our results for absorption of a kinetically mixed dark photon DM are shown in the right panel of Fig.~\ref{fig:bounds}.  
Similarly to the left panel, the blue shaded region indicates the new bounds we extract from the measurements of Ref.~\cite{Connolly:2023gww}, where line thickness corresponds to different values of $s$, with the highest sensitivity reached for $s=10$ Hz. The bound obtained using the data of Ref.~\cite{Riste:2013zqw}, shown by the dashed orange curve, is weaker by ${\cal O}(1)$ compared to the bound obtained using the more recent measurement of Ref.~\cite{Connolly:2023gww}, similarly to the DM scattering case. 
The gray shaded region shows existing terrestrial constraints assuming halo DM dark photon \cite{EDELWEISS:2019vjv, EDELWEISS:2022ktt, DarkSide-50:2022qzh, Franco:2023sjx, SuperCDMS:2020aus, SuperCDMS:2023sql, CRESST:2019jnq, SENSEI:2023zdf, PandaX:2023xgl, LUX:2018akb, Essig:2019xkx, QROCODILE:2024zmg, Barak:2020fql, Amaral:2020ryn, Aguilar-Arevalo:2019wdi, Essig:2017kqs, Agnes:2018oej, XENON:2019gfn, An:2014twa, Agnese:2018col, Aguilar-Arevalo:2019wdi, Arnaud:2020svb, FUNKExperiment:2020ofv, Barak:2020fql, Chao:2024owf}, while the yellow-shaded band represents complementary stellar constraints on absorption \cite{An:2013yua, An:2020bxd}. 
We also show laboratory constraints on conversion of photons to dark photons \cite{Bahre:2013ywa} as a light green band. 
We find that the transmon bound we place surpasses other terrestrial constraints for dark photon masses below $0.1\,{\rm eV}$, and is competitive with conversion bounds. 
A possible future measurement of a lower $x_{\rm qp}$, {\it e.g.} $x_{\rm qp}=10^{-11}$, could provide leading laboratory constraints on the dark photon DM mass in the range $\sim 0.01 - 0.1 \, {\rm eV}$.

\section{Outlook}
In this work, we showed how coherence time measurements~\cite{Riste:2013zqw,Connolly:2023gww} from transmons already place world-leading lab-based constraints on spin-independent DM-electron scattering, and competitive bounds on kinetically-mixed dark photon absorption on electrons.
Given the rapid advancements and large-scale funding of transmon-qubit-based quantum computing efforts of recent years, we expect future devices to probe even smaller DM-electron couplings.
Here, we placed conservative bounds on DM interactions with electrons, corresponding to the assumption that the power injection from DM alone does not exceed the observed amount.
Improvements in experimental methods, isolation from noise sources, modeling of these quantum systems and characterization of their noise would all provide improved sensitivity to smaller DM couplings.

Assuming the transmon parameters $r$ and $s$ (see Eq.~\eqref{eq:qp master equation}) remain of similar order to their current values, we expect future measurements of smaller residual QP fractions $x_{\rm qp}$ to be essentially sensitive only to the trapping rate, since the QP recombination rate $r\,x_{\rm qp}^2$ is quadratically suppressed by the QP fraction whereas the QP trapping rate $s \, x_{\rm qp}$ is only linearly suppressed. As a result, we anticipate future measurements to yield DM limits scaling according to Eq.~\eqref{eq:trapping bound}, such that $\bar \sigma_e \propto s \, x_{\rm qp}$ for DM-electron scattering limits and $\kappa\propto \sqrt{s \, x_{\rm qp}}$ for kinetically mixed dark photon absorption limits.  
Reduced excess QP populations and lower trapping rates 
in future transmon fabrication and measurements will translate directly into significant improvement on DM limits. 
(For a recent complementary DM proposal using qubits as sensors coupled to led to a sapphire target, aiming to resolve individual scattering events as QP bursts, see Ref.~\cite{Li:2025hqk}.)

Interestingly, we note that transmons offer a potential avenue for directional detection, where a daily modulation of the DM signal could arise from 
anisotropies in the response of the transmon due to its thin film morphology, as well as via Earth shielding~\cite{DAMIC-M:2023hgj,SENSEI:2025qvp}.
Probing such a signal would require a continuous time measurement of the qubit over a day. 
The practicality of directional detection of DM with transmons is left for future work.

\textbf{Acknowledgments.} 
We thank Valla Fatemi for useful discussions and comments on the manuscript.
The work of Y.H. is supported by the Israel Science Foundation (grant No. 1818/22) and by the Binational Science Foundation (grant No. 2022287). Y.H., M.K., A.L. and R.O. are supported by an ERC STG grant (``Light-Dark,'' grant No. 101040019). 
M.K., A.L. and R.O. are each grateful to the Azrieli Foundation for the award of an Azrieli Fellowship. 
The work of M.K. is also supported by the BSF grant No.
2020220. 
R.O. is also supported by BSF Travel Grant No. 3083000028 and the Milner Fellowship. 
Y.H, M.K., A.L and R.O. thank Cornell University for their gracious hospitality. N.K. is supported by the US Department of Energy Early Career Research Program (ECRP) under FWP 100872. This project has received funding from the European Research Council (ERC) under the European Union’s Horizon Europe research and innovation programme (grant agreement No. 101040019).  Views and opinions expressed are however those of the author(s) only and do not necessarily reflect those of the European Union. The European Union cannot be held responsible for them.

\end{document}